# Spiral growth, two-dimensional nucleation, and the Ehrlich-Schwoebel effect

**Joachim Krug**

*Institut für Theoretische Physik, Universität zu Köln,*
*Zülpicher Strasse 77, 50973 Köln, Germany*
krug@thp.uni-koeln.de

**Abstract.** Frank's prediction of the spiral growth mode in 1949 defined a pivotal moment in the history of crystal growth. In recent decades the topic has received less attention, and instead we have seen a resurrection of two-dimensional nucleation theory in the context of growth experiments on defect-free homoepitaxial thin films. In particular, the key role of interlayer transport controlled by step edge barriers of the Ehrlich-Schwoebel type in shaping the morphology of multilayer films has been increasingly recognized. After a brief review of the classical theory, this paper reports on a recent study of spiral growth in the presence of step edge barriers. Our key observation is that step edge barriers lead to unconventionally shaped spiral hillocks that display the same characteristic ever-steepening height profiles as wedding cakes formed during growth by two-dimensional nucleation. This prediction was verified experimentally by inducing screw dislocations through ion bombardment of the Pt(111) surface, thus creating a homoepitaxial growth system on which spiral hillocks and wedding cakes coexist.

> *Wir mussten annehmen, es walte in der Vegetation eine allgemeine Spiraltendenz, wodurch, in Verbindung mit dem vertikalen Streben, aller Bau, jede Bildung der Pflanzen, nach dem Gesetze der Metamorphose, vollbracht wird.*  J.W. Goethe (1831)

## 1. Background

In 1947 the South African physicist Charles Frank joined the physics department at Bristol and was asked by the department head, Nevill Mott, to prepare a lecture course on crystal growth [1]. Having no previous expertise in the subject, he immersed himself into Volmer's classic book on nucleation theory [2] and noticed (in discussions with W. Burton and N. Cabrera) that the theoretical predictions for crystal growth rates from the vapor phase were in fact grossly inconsistent with experimental data. He then made the ingenious suggestion that the growth of real crystals is facilitated by screw dislocations which intersect the crystal surface and thereby provide a constant source of steps to which the growth units can attach, without the need of creating new steps by two-dimensional nucleation [3]. Frank's prediction was soon verified by direct visualization of growth spirals [1], and the landmark article of Burton, Cabrera and Frank (BCF) [4] remains one of the most cited papers in the field.

Following the invention of the scanning tunneling microscope (STM), the past two decades have seen a resurgence of interest in the fundamentals of crystal growth kinetics. As much of the experimental work has been concerned with essentially defect-free homoepitaxial growth systems, attention has turned away from spiral growth towards the classic topic of two-dimensional nucleation theory and its extension to multilayer growth [5,6]. In particular, the key importance of additional energy barriers at step edges controlling the material transport between different layers of the growing film has been increasingly recognized. The existence of such barriers was postulated in 1966 by Ehrlich and Hudda in a field ion microscopy study of tungsten atoms on the tungsten surface [7], and some of the consequences for the morphological stability of growing stepped surfaces were soon afterwards spelled out by Schwoebel and Shipsey [8]. The appropriately named *Ehrlich-Schwoebel (ES) effect* is now a central paradigm in crystal growth theory, and novel ramifications are constantly being discovered.

In this paper we will let Burton, Cabrera and Frank meet with Ehrlich and Schwoebel; that is, we will consider the consequences of the ES effect on spiral growth. To set the stage, the next section contains a brief review of spiral growth theory. In Section 3 we then compare the growth of spiral hillocks to the well-known phenomenon of mound formation due to the ES effect [5,6] and present some results of a recent combined experimental and computational study of spiral growth in the presence of ES barriers [9].

## 2. Theories of spiral growth

The BCF theory of spiral growth [4] is based on expressing the normal velocity $v_n$ of the step in terms of its local curvature $\kappa$ according to $v_n = v_\infty (1 - \kappa\, r_c)$, where $v_\infty$ is the velocity of a straight step and $r_c$ is the radius of the critical two-dimensional nucleus. At the spiral core the step velocity vanishes, and hence the radius of curvature at the center of the spiral is just $r_c$. The ansatz of a stationary spiral rotating at a fixed angular velocity leads to an ordinary differential equation for the spiral shape parametrized by the rotation angle $\varphi(r)$ around the core. As the curvature at the origin is given by $\kappa(0)=2\,\varphi'(0)$, the simplest assumption of an Archimedean spiral implies that $\varphi(r)=2r/r_c$, and hence the step spacing between subsequent turns is $4\pi\,r_c$. A more careful analysis yields a somewhat larger step spacing $l\approx 19\,r_c$ far away from the core [10]. Thus the spiral hillock forming around the screw dislocation is steeper near the center, an effect that can be quantified by the *tapering factor* $T=l\,\kappa(0)/4\pi$ (for an Archimedean spiral $T=1$) [11].

This *local* description is valid as long as the diffusion length $l_D\equiv\sqrt{D\tau}$ that an atom travels along the surface before desorbing is small compared to $l$ and $r_c$ (here $D$ denotes the surface diffusion coefficient and $1/\tau$ the desorption rate). Otherwise, the different parts of the spiral are coupled by the diffusion field and the shape of the spiral is determined by a *nonlocal* moving boundary value problem in a complicated geometry. Qualitatively, the most important effect of surface diffusion is to reduce the supersaturation near the spiral core, because adatoms are captured by the surrounding turn of the spiral. This amounts effectively to an increase of $r_c$ and hence to a flattening of the spiral hillock.

For a quantititave treatment of this so called *back-force effect* Surek, Hirth and Pound (SHP) approximated the spiral geometry by a stack of concentric circular islands (compare to Fig. 1) [12,13]. The two step configurations are related by identifying the radii of the first (innermost) and second islands, $r_1$ and $r_2$, to the radius of curvature at the spiral core ($\varphi=0$) and after one spiral turn ($\varphi=2\pi$), respectively. The growth rate of the concentric circles is obtained by solving the stationary BCF equation for the adatom concentration on the terraces, and a new island (corresponding to a new turn of the spiral) of size $r_1$ is added at the center of the stack when the radius of the current innermost island reaches $r_2$. The values of $r_1$ and $r_2$ are fixed through two relations. The first identifies $r_1$ as the critical radius of a two-dimensional nucleus growing in the diffusion field on top of a circular island of radius $r_2$, while the second relation reflects the geometry of the spiral; e.g., assuming the spiral to be Archimedean one has $r_2/r_1=4\pi$.

The predictions obtained from the approximate SHP approach are confirmed (at least semiquantitatively) by a more rigorous integral equation formalism that takes into account the proper spiral geometry [11,14] as well as by phase field simulations [15]. The different kinetic regimes can be distinguished by the resulting scaling of the asymptotic step spacing $l$ with the net deposition flux $f$. In the local (BCF) regime $l \propto r_c \propto f^{-1}$, whereas in the nonlocal regimes $l \propto f^{-1/3}$ or $l \propto f^{-1/2}$ depending on whether the attachment of adatoms to steps is fast or slow compared to the diffusion on the terraces. Behavior compatible with the attachment-limited regime ($l \propto f^{-1/2}$) has recently been observed for growth and sublimation spirals on Si(111) [16].

In the discussion so far it has been assumed that the spiral step can be well described as a smoothly curved object. However, in many systems with strong crystalline anisotropy (e.g. [17]) spirals consist of straight step segments separated by corners, and the notion of a local step curvature κ and a critical radius $r_c$ cannot be applied. The scale of such a *polygonized* spiral is instead set by the *critical length* $l_c$ that a newly created step segment emerging from the core has to reach before it can start to grow normal to itself, i.e. before a kink has been formed by one-dimensional nucleation [18]. While the kinematics of polygonized spirals has been worked out in detail [19], the coupling to the diffusion field does not appear to have been considered so far.

**3. Spiral hillocks and wedding cakes**

The common feature of the theoretical approaches reviewed in the preceding section is that the resulting spirals are always close to Archimedean, in the sense that the tapering factor $T$ is close to unity [11,14]; correspondingly, the spiral hillocks are essentially conical mounds of constant slope. We will now argue that the shape of the hillock changes dramatically in the presence of a strong Ehrlich-Schwoebel effect. To see why this is so, we recall the similarity between a spiral and a stack of concentric islands, which (for obvious reasons) we will refer to in the following as a *wedding cake* (Fig. 1). The success of the SHP approximation clearly relies on the fact that the diffusion geometry encountered by an adatom incident on the hillside of the structure, a few turns (or island boundaries) away from the center, is indistinguishable in the two cases. In the presence of a strong ES effect, such an atom will attach with high probability to the ascending step bordering the terrace.

As was first pointed out by Jacques Villain [20], the migration of the atom is therefore effectively *biased* in the direction of increasing height, an mechanism which destabilizes the planar surface and gives rise to the formation of mounds [5].

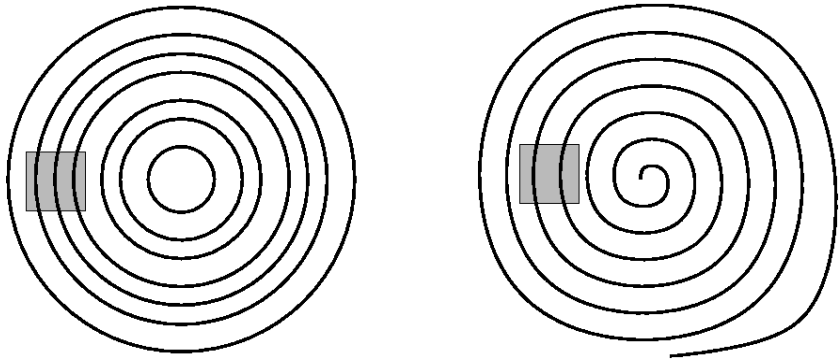

**Fig. 1.** Schematic comparison of a wedding cake (left) and a spiral hillock (right). The diffusion geometry encountered by an adatom on the hillside of the structure (shaded area) is indistinguishable in the two cases.

Villain's argument is macroscopic, and as such it applies equally well to both geometries depicted in Fig. 1. Detailed studies of the shape evolution of mounds formed on metal surfaces (which often display a strong ES-barrier) have been based on a wedding cake model [21]-[23] that is essentially identical to the SHP model described in the preceding section, with two modifications necessitated by the fact that the growth process of interest proceeds far from equilibrium. First, new islands created at the center of the stack represent a microscopic cluster and hence have zero radius ( $r_1 = 0$ ); second, the critical size $r_2$ of the current top terrace at which the new island appears is determined by kinetic nucleation theory [24] rather than by thermodynamic considerations. A similar model was studied already thirty years ago in the context of pit formation during sputtering [25].

When the ES barrier is large, in the sense that an atom impinging on the hillside of the mound always attaches to the ascending step, the equations governing the areas of the islands that constitute the wedding cake become linear [21,26,27] and a simple analytic solution is possible in the limit of large deposited coverage [23]: The fractional coverage

$\theta(h)$ at (discrete) layer height *h*, which is proportional to the area of the corresponding island in the wedding cake, is given by the expression

$$\theta(h) = 1 - C\{1 + \mathrm{erf}[(h-\Theta)/\sqrt{\Theta}]\}, \qquad (1)$$

where $\Theta$ denotes the total deposited coverage, erf(*x*) is the error function and the constant *C* is set by the condition that the sum over $\theta(h)$ up to the maximal height $h_{max}$ of the top terrace should equal $\Theta$. At the maximal height the fractional coverage reaches the value $\theta(h_{max}) = \theta_c$ corresponding to the critical island radius $r_2$ at which nucleation occurs. Equation (1) provides an accurate description of the shape of wedding cakes grown on the Pt(111) surface [28], and the value of $\theta_c$ derived from the experiments yields an estimate of the ES barrier [5,24].

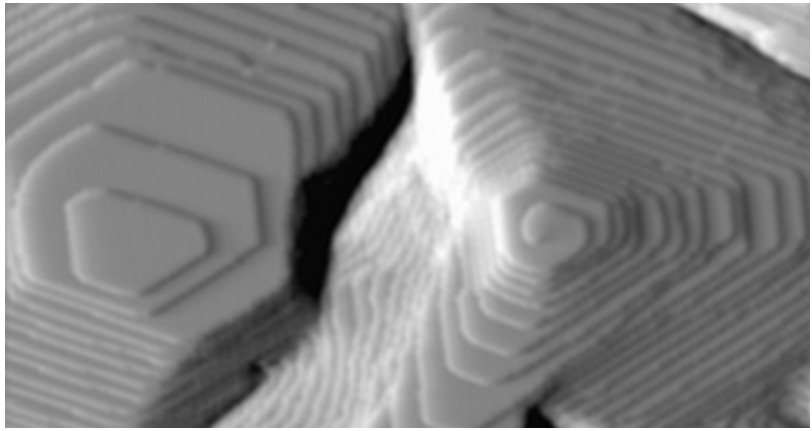

**Fig. 2.** Scanning tunneling microscopy image of a Pt(111) surface displaying a wedding cake (left) and a spiral hillock (right). The latter is several monolayers higher than the former. The horizontal size of the image corresponds to 91 nm. Courtesy of A. Redinger, O. Ricken and T. Michely.

It is clear from the preceding discussion that the coverage profile (1) should be realized also in the shape of spiral hillocks grown in the presence of ES barriers, possibly with some modification in the top part of the structure where the physical processes near the core of the spiral become relevant. Equation (1) differs from the conical spiral hillocks predicted by the classical theories in two important respects. First, the slope depends on the layer height *h*, displaying a minimum at $h = \Theta$; second, the hillock shape does not become stationary, but rather *steepens*

*indefinitely* with the typical step spacing decreasing as $l \propto \Theta^{-1/2}$. This prediction was verified by phase field simulations [9,29] in which the ES effect was implemented through a suitably chosen mobility function for the phase field [30].

An experimental system in which wedding cakes and spiral hillocks can be compared under identical conditions was created by inducing screw dislocations in a Pt(111) sample by bombardment with a 4.5 keV $He^+$ ion beam prior to deposition [9,31]. The resulting growth morphology displays a mixture of wedding cakes and spiral hillocks of very similar shapes, but, surprisingly, the spiral hillocks are distinctly higher (Fig. 2). To understand this observation, we note that, within the model (1), the height of a mound is determined by the lateral extent of the top terrace (the critical coverage $\theta_c$ or radius $r_2$ at which a new layer is added): The smaller the top terrace, the higher the mound. As the spirals grown on the Pt(111) surface are strongly polygonized, the physics of the core region is determind by one-dimensional nucleation (see Sect.2) and the size of the top terrace is set by the critical length $l_c$ of the first step segment emanating from the core. The evaluation of several growth spirals yielded the estimate $l_c \approx 23 \pm 6$ Å, which is much smaller than the size of the top terraces of wedding cakes governed by two-dimensional nucleation. For the wedding cakes to reach the same height as the spiral hillocks, the step edge barrier would have to be increased from about 0.2 eV to about 0.3 eV [9].

## 4. Conclusions

In this paper I have discussed the interplay of screw dislocations and step edge barriers, which constitute two of the most important factors shaping the morphology of growing crystalline films. In a broader perspective, work along these lines may provide a basis for the exploration of kinetic growth instabilities combined with dislocation engineering as a tool for the lateral and vertical patterning of thin film surfaces.

**Acknowledgments**

The paper is based on joint work with Philipp Kuhn, Thomas Michely, Andreas Rätz, Alex Redinger, Oliver Ricken and Axel Voigt.